\documentclass[preprint,showpacs,preprintnumbers,amsmath,amssymb]{revtex4}
\usepackage{graphicx,epsfig,dcolumn,bm,epic,eepic,float}% Include figure files
\usepackage{amsmath}
\usepackage{latexsym}
\usepackage{color}
\usepackage{makeidx,shortvrb,latexsym}
\begin{document}
\unitlength 1 cm
\newcommand{\be}{\begin{equation}}
\newcommand{\ee}{\end{equation}}
\newcommand{\bearr}{\begin{eqnarray}}
\newcommand{\eearr}{\end{eqnarray}}
\newcommand{\nn}{\nonumber}
\newcommand{\vk}{\vec k}
\newcommand{\vp}{\vec p}
\newcommand{\vq}{\vec q}
\newcommand{\vkp}{\vec {k'}}
\newcommand{\vpp}{\vec {p'}}
\newcommand{\vqp}{\vec {q'}}
\newcommand{\bk}{{\bf k}}
\newcommand{\bp}{{\bf p}}
\newcommand{\bq}{{\bf q}}
\newcommand{\br}{{\bf r}}
\newcommand{\bR}{{\bf R}}
\newcommand{\up}{\uparrow}
\newcommand{\down}{\downarrow}
\newcommand{\fns}{\footnotesize}
\newcommand{\ns}{\normalsize}
\newcommand{\cdag}{c^{\dagger}}
\title {Extracting the parton distribution functions evolution equations using the stochastic modeling in the non-equilibrium statistical mechanics}
\author{N. Olanj$^\dag$}\altaffiliation {Corresponding author, Email :   {$n\_olanj@basu.ac.ir$}
, Tel:+98-81-38381601}
\author{E. Moradi$^\dag$}
\author{M. Modarres$^\ddag$}
\affiliation{$^\dag$Physics Department, Faculty of Science, Bu-Ali Sina University, 65178, Hamedan, Iran} 
\affiliation{$^\ddag$Physics Department, University of Tehran, 1439955961, Tehran, Iran.}
\begin{abstract}
In this paper, using the stochastic modeling of the non-equilibrium statistical mechanics in the momentum space,
 the evolution equations of the parton distribution functions ($PDF$) usually used in the hadrons phenomenology are generated. 
These stochastic modeling $PDF$ evolution equations are the same as those of the $Dokshitzer$-$Gribov$-$Lipatov$-$Altarelli$-$Parisi$ ($DGLAP$) ones, but they can be obtained by a more simplistic mathematical procedure based on the non-equilibrium statistical mechanics and the theory of Markov processes.
\end{abstract}
\maketitle
\section{Introduction}
The integrated parton distribution functions ($PDF$) are the main objects of phenomenological computations
 in the high energy collisions of particle physics. These $PDF$ are usually obtained by using the experimental data
  via the parameterizations procedures which are constrained by the sum rules and a few theoretical assumptions, e.g., see the reference \cite{1modarres}. 
\\Today, many of the high-energy particle physics laboratories, including the $LHC$, use the $PDF$ to describe and analyze their extracted data from the deep inelastic scatterings, e.g., in the various $LHC$ detectors such as $ATLAS$, $CMS$, $LHCb$ and $ALICE$ \cite{2modarres, 3modarres, 4modarres, 5modarres}. 
 \\ Because of the importance of this subject, the $Dokshitzer$-$Gribov$-$Lipatov$-$Altarelli$-$Parisi$ ($DGLAP$) obtained the evolution equations for the $PDF$ in terms of $Q^2$ (the hard scale of probe) in the lowest order level of the quantum chromodynamics ($QCD$). The $DGLAP$ evolution 
 equations are generated \cite{1a,1b,1c,1d} by performing the Mellin transform techniques starting from the renormalization group equations, as follows: 
\begin{eqnarray}
\frac{\partial q(x, Q^{2})}{\partial Ln Q^2} &=&\frac{\alpha_s({Q}^2)}{2\pi}\int_x^{1}\frac{dz}{z}\Bigg[P_{qq}(z)q\left(\frac{x}{z}, {Q}^2 \right)+P_{qg}(z) g\left(\frac{x}{z}, {Q}^2 \right)\Bigg],
\nonumber
\\\frac{\partial g(x, Q^{2})}{\partial Ln Q^2}&=&\frac{\alpha_s({Q}^2)}{2\pi}\int_x^{1}\frac{dz}{z}\Bigg[P_{gg}(z)g\left(\frac{x}{z}, {Q}^2 \right)+P_{gq}(z)\sum_q q\left(\frac{x}{z}, {Q}^2 \right)\Bigg],
\label{eq:DGLAP}
\end{eqnarray}
where $q(x, Q^2)$ and $g(x, Q^2)$ are the quark or anti-quark and the gluon distribution functions, 
respectively. $P_{aa'}(z)$ $(a, a'=q, \overline{q}, g)$, the probability of emitting parton $a$ with fraction $z$ of the longitudinal momentum of the parton $a'$, are the leading order ($LO$) splitting functions. $Q ^2$ is the hard scale of the scattering that comes from the virtuality of the space like exchanged photon ($q^2$ $\equiv$  $q^\mu q_\mu $ = $-Q^2$), and $x$ is a fraction of the longitudinal momentum of the parent hadron (the Bjorken variable). In the other words, $q_i(x, Q^2)$ ($g(x, Q^2)$) is the probability of the finding the quark or anti-quark of type $i$ (gluon) with the fraction $x$ of longitudinal momentum of the parent hadron inside the hadron by  probing photon with the energy scale $Q^2$. 
It is worth noting that  $Kimber$, $Martin$ and $Ryskin$ ($KMR$)  \cite{18, 19} modified the standard $DGLAP$ equations due to the separation of virtual and real parts of the evolutions. Therefore, the $DGLAP$ equation is rewritten as follows: 
\begin{eqnarray}
\frac{\partial a(x, Q^{2})}{\partial Ln Q^2} =\frac{\alpha_s({Q}^2)}{2\pi}\sum_{a'= q,g}\Bigg[\int_0^{1-\Delta}dz P_{aa'}(z)a'\left(\frac{x}{z}, {Q}^2 \right)-a\left(x, {Q}^2 \right)\int_0^{1-\Delta}dz P_{a'a}(z) \Bigg],
\label{eq:KMR}
\end{eqnarray}
where $a(x, Q^2)$ denotes  $xg(x, Q^2)$ or $ xq(x, Q^2)$ and $\Delta$ is a cut-off to prevent $z = 1$
 singularities in the splitting functions which arises from the soft gluon emissions. 
\\ Also, the $Q^2$ evolution of the $PDF$ determined by $QCD$ were used to investigate the
  deep inelastic structure function ($F_2(x, Q^2)$) of the proton \cite{halzen}. The resulting
   evolution equation in the leading order level is the same as the $DGLAP$ evolution equations. 
In the present phenomenological studies, it is observed that the $PDF$ with the increasing the hard scale $Q^2$ of the probe never reaches a stable state. So one can conclude that the prominent role of non-equilibrium statistical mechanics 
in describing the evolution of the $PDF$ in the hadron is vital. Then it would be interesting to show that the above evolution equation can be obtained, using some simplistic  stochastic modeling based on the non-equilibrium statistical mechanics.
\\Therefore, in the reference \cite{Nayak}, Nayak studied the splitting functions in the non-equilibrium 
$QCD$ at the leading order level and Buccella et al, in the reference \cite{Buccella}, 
obtained the low $Q^2$ boundary conditions for $DGLAP$ equations by using the quantum statistical mechanics. 
\\So, considering that,  the process of the parton $Q^2$ evolution in the leading order 
level has completely stochastic nature, and depends only on a previous step, i.e., Markov processes \cite{6modarres}, 
in this work, we intend to study the $Q^2$ evolution of the $PDF$ in the non-equilibrium $QCD$ by using the stochastic modeling. 
It is shown that, these evolution equations are the same as those of $DGLAP$, but they can be generated  by a much "simpler" mathematical procedure, using the stochastic modeling of the non-equilibrium statistical mechanics and the theory of Markov processes. 
\\The parton $Q^2$ evolution equations derived from this procedure are expressed by using the master
 equations (the evolution equation of stochastic variables of Markov processes) of statistical mechanics of systems far from equilibrium in the momentum space. 
  This master equation is equivalent to the master equation in the position space
   for the single-particle systems with the multi-species reaction diffusion processes on 
   the one-dimensional finite continuous lattice. It should be noted that the master
    equation in the position space (momentum space) is an integro-differential equation 
    governing the time (energy) evolution of the probability \cite{Reichl}. 
\\So the paper is organized as follows: the Sect. $II$ contains an overview of the master equation of Markov processes. Sect. $III$ is devoted to study the $Q^2$ evolution 
of the $PDF$ inside the hadron by using stochastic modeling in the non-equilibrium statistical mechanics and the theory of Markov processes.
 Finally, Sect. $IV$ contains results, discussions and conclusions. 
 \section{An overview of derivation of the master equation of Markov processes}
The master equation is the evolution equation of stochastic variables of Markov processes. It should be noted that Markov processes are stochastic processes that depend only on a limited history of evolution. The master equation is one of the most important statistical physics equations due to its many applications in various sciences areas.  In this section, we take a brief look at how to derive the master equation of non-equilibrium statistical mechanics in the position space \cite{Reichl}.
\\First, we define the notations for the probability density,  the joint probability density and the conditional probability density  for the stochastic variable $Y$, respectively, as follows:
\\(i) $P_1(y_1, t_1)$: the probability density that $Y$ has a value $y_1$ at time $t_1$,
\\(ii) $P_2(y_1, t_1; y_2, t_2)$: the joint probability density that $Y$ has a value $y_1$ at time $t_1$ and a value $y_2$ at time $t_2$,
\\(iii) $P_{1\mid1}(y_1, t_1\mid y_2, t_2)$: the conditional probability density that $Y$ has a value $y_2$ at time $t_2$ given that it had a value $y_1$ at time $t_1$, so that the probability density is normalized as:
 \begin{eqnarray}
\int dy_1 P_{1}(y_1, t_1) = 1,
\label{eq:norm}
\end{eqnarray}
and the joint probability density can be reduced, as follows:
 \begin{eqnarray}
\int dy_2  P_2(y_1, t_1; y_2, t_2) = P_{1}(y_1, t_1),
\label{eq:reduced}
\end{eqnarray}
also, the conditional probability density satisfies the following equation:
 \begin{eqnarray}
P_1(y_1, t_1)P_{1\mid 1}(y_1, t_1\mid y_2, t_2) = P_2(y_1, t_1; y_2, t_2).
\label{eq:identity}
\end{eqnarray}
Finally, by combining equations (\ref{eq:reduced}) and (\ref{eq:identity}), we arrive at the following relation between probability densities at different times:
 \begin{eqnarray}
P_1(y_2, t_2) = \int dy_1 P_1(y_1, t_1)P_{1\mid 1}(y_1, t_1\mid y_2, t_2).
\label{eq:identity1}
\end{eqnarray}
It should be noted that equation (\ref{eq:identity1}) is obtained assuming that the stochastic variable $Y$ is continuous. If we consider the stochastic variable $Y$ to be discrete, the equation (\ref{eq:identity1}) simply becomes as following:
\begin{eqnarray}
P_1(n, t+\Delta t) =\sum_{m=1}^{M}P_1(m, t)P_{1\mid 1}(m, t\mid n, t+\Delta t),
\label{eq:identity2}
\end{eqnarray}
where $M$ is the total number of states. Using the derivative definition, the differential equation for $P_1( n, t)$ can be obtained from the equation (\ref{eq:identity2}):
\begin{eqnarray}
\frac{\partial P_1(n, t)}{\partial t} &=& \lim_{\Delta t\rightarrow 0}\Bigg(\frac{P_1(n, t+\Delta t)-P_1(n, t)}{\Delta t}\Bigg)\nonumber
\\ &=& \lim_{\Delta t\rightarrow 0} \frac{1}{\Delta t} \sum_{m=1}^{M}P_1(m, t)\Bigg(P_{1\mid 1}(m, t\mid n, t+\Delta t)-\delta_{mn}\Bigg).
\label{eq:identity3}
\end{eqnarray}
By expanding the transition possibility $P_{1\mid 1}(m, t\mid n, t+\Delta t)$ in a power series in $\Delta t$ and keeping only the lowest order term, the transition possibility becomes:
\begin{eqnarray}
P_{1\mid 1}(m, t\mid n, t+\Delta t) = \delta_{mn}\Bigg[1-\Delta t\sum_{\ell=1}^{M}\omega_{m,\ell}(t)\Bigg]+\omega_{m,n}(t)\Delta t,
\label{eq:identity4}
\end{eqnarray}
where $\omega_{m,n}(t)$ is the transition probability rate. Therefore, the first term on the right of the equation (\ref{eq:identity4}) ($\delta_{mn}[1-\Delta t\sum_{\ell=1}^{M}\omega_{m,\ell}(t)]$) is the probability that no transition occurs at time interval $t$ to 
$t+\Delta t$, and the second term ($\omega_{m,n}(t)\Delta t$) implies the probability of a transition from state $m$ to state $n$ at time interval $t$ to $t+\Delta t$.
\\By combining the equation (\ref{eq:identity3}) and the equation (\ref{eq:identity4}), and given that we have only one transition (Markov processes) in the equation (\ref{eq:identity4}), we obtain the master equation of Markov processes for the discrete stochastic variable, as follows:
\begin{eqnarray}
\frac{\partial P_1(n, t)}{\partial t} = \sum_{m=1}^{M}[P_1(m, t)\omega_{m,n}(t)-P_1(n, t)\omega_{n,m}(t)].
\label{eq:identity5}
\end{eqnarray}
The master equation (the equation (\ref{eq:identity5})) describe the time evolution of the probability $P_1(n, t)$ due to transitions into the state $n$ from all others states (the first $M$ terms on the right which are called source terms) and due to transitions out of state $n$ into all others states (the second $M$ terms on the right which are called sink terms). Assuming the stochastic variable is continuous, the equation (\ref{eq:identity5}) for the infinite continuous stochastic variable $X$ is written as follows:
\begin{eqnarray}
\frac{\partial P_1(x, t)}{\partial t} =\int_{-\infty}^{+\infty}dx^{'} \Bigg[P_{1}(x^{'}, t)\omega(x^{'}\mid x, t)-
P_{1}(x, t)\omega(x\mid x^{'}, t)\Bigg],
\label{eq:identity6}
\end{eqnarray}
where $\omega(x^{'}\mid x, t)$ is the transition rate. Now considering that the value of the stochastic variable $X$ changes from $x$ to $x^{'}=x+y$ at a transition and by introducing the notation $\tau(x, y, t) = \omega(x\mid x+y, t)$, the master equation can be rewritten as follows:
\begin{eqnarray}
\frac{\partial P_1(x, t)}{\partial t} = \int_{-\infty}^{+\infty}dy \Bigg(P_{1}(x-y, t)\tau(x-y, y, t)-
P_{1}(x, t)\tau(x, y, t)\Bigg).
\label{eq:Master0}
\end{eqnarray}
Finally, the generalization of the master equation (\ref{eq:Master0}) in the position space, i.e., the stochastic variable $X$ represents the position of the particle, for the single-particle systems with the $p$-species reaction diffusion processes is as follows \cite{Reichl, ahmadi}: 
\begin{eqnarray}
\frac{\partial P_i(x, t)}{\partial t} =\sum_{j=1}^{p}\Bigg[\int_{-\infty}^{+\infty}dy \Bigg(P_{j}(x-y, t)\tau_{ij}(x-y, y, t)-
P_{i}(x, t)\tau_{ji}(x, y, t)\Bigg)\Bigg],
\label{eq:Master}
\end{eqnarray}
where $P_i(x, t)$ is the probability of finding at time $t$ the particle of type $i$  at the point $x$ and  $\tau_{ji}(x, y, t)$
 is the process rate that changes the position of the particle from $x$ to $x+y$ and the particle type from $i$ to $j$ at the time $t$. 
 It should be noted that the first $p$ terms are the sources of $P_i(x, t)$ and the second $p$ terms are the sinks of it.  
\section{ The study of the parton $Q^2$ evolution equations in the non- equilibrium $QCD$ by using stochastic modeling }
Based on the master equation (equation (\ref{eq:Master})),  probing the structure of the hadron, by a virtual photon with virtuality $Q^2$ that coming from the projectile, e.g., electron, in the deep inelastic scattering, can be viewed as a portion of the parton evolution chain demonstrated in the figure 1 \cite{tkimber}.
 \begin{figure}[ht] 
\includegraphics[width=160mm]{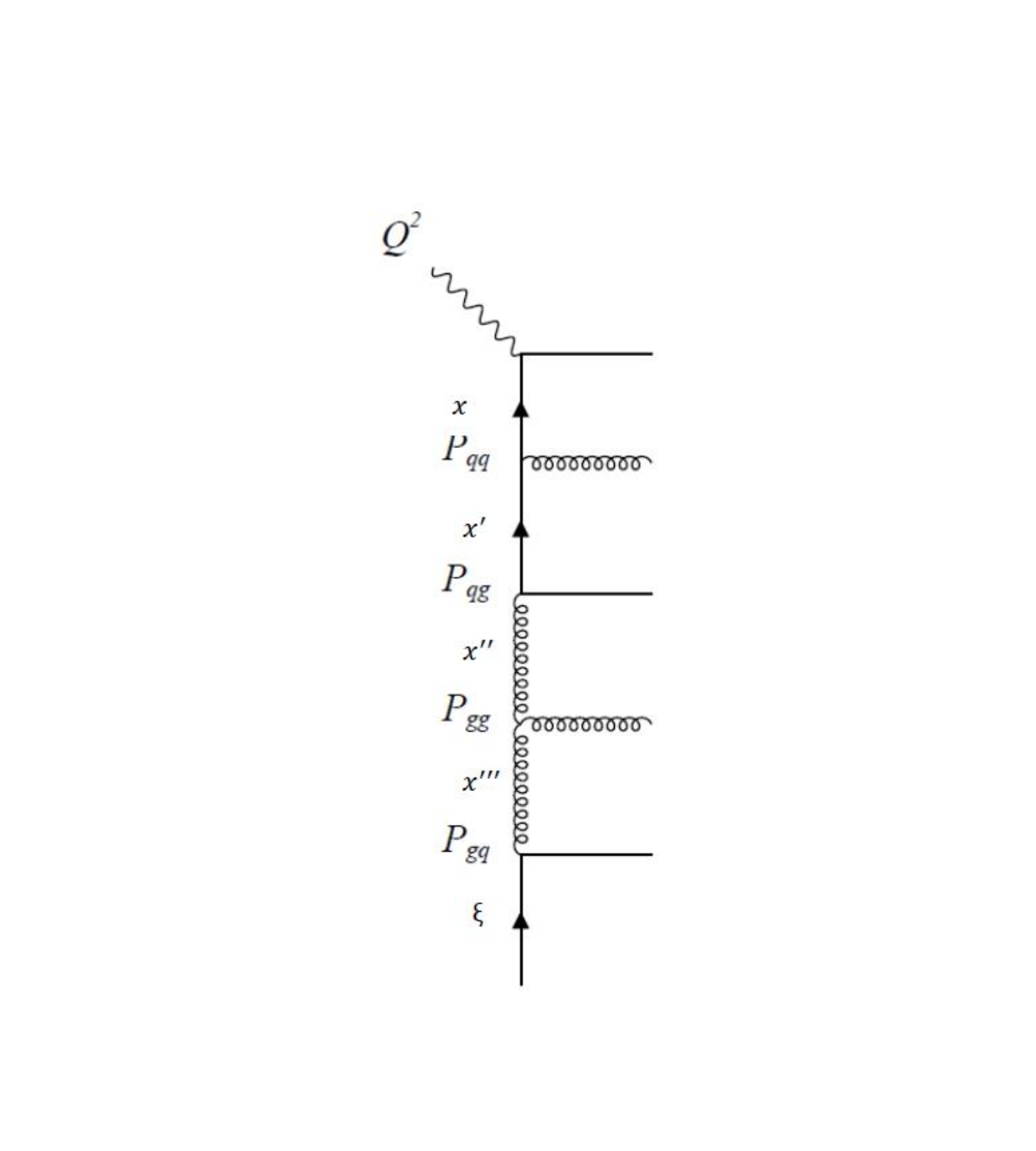}
\caption{A portion of the evolution chain (the different types of parton splitting in the evolution chain at the scale $Q^2$ \cite{tkimber}). }
\label{fig:1}
\end{figure} 
The splitting process at each stage of the evolution is a completely stochastic process and in each splitting, the fraction of the longitudinal momentum decreases, so:
\begin{eqnarray}
x<x'<x''<x'''<...<\xi .
\label{eq:x}
\end{eqnarray}
\\According to this diagram, a parton on the one-dimensional finite ($0<x<1$) continuous lattice 
(longitudinal momentum axis) changes its longitudinal momentum and type (convert quark to gluon and vice versa- flavour of quark) during the reaction diffusion
 processes to the scale $Q^2$. 
\\According to the experimental evidence, the probability of the finding a parton type $i$ with the fraction 
$x$ of longitudinal momentum (the Bjorken variable) of the parent hadron inside the hadron with the increase in the energy
 of the virtual photon ($Q^2$) never gets into a stable state \cite{19modarres}. In other words, the parton structure of the
  hadrons will never be saturated with increasing the energy of the virtual photon ($Q^2$). Therefore, 
  the process of the parton $Q^2$ evolution follows the non-equilibrium statistical mechanics and can satisfy the master equation (equation (\ref{eq:Master})) conditions. 
\\Of course, it should be noted that, at the leading order level of the $QCD$, the probability of finding a parton type $i$ that carrying the fraction $x$ of longitudinal momentum of the parent hadron in the hard scale $Q^2$ of the probe, depends only on a previous step and  not on the history of evolution, i.e., Markov processes. 
\\Then, the processes performed by the partons, corresponding to transitions between different positions in the master equation in the position space (\ref{eq:Master}) with the transition rate $\tau_{ji}(x, y, t)$, at the leading order level inside the hadron can be expressed as follows: 
\begin{eqnarray}
1)\ \  &g(\frac{x}{z}, Q^2)&\longrightarrow q_{i'}(x, Q^2) \ \  \text{  with rate  }  \ \   \frac{\alpha_s(Q^2)}{2\pi}P_{qg}(z),\nonumber
\\2)\ \   &q_{i}(\frac{x}{z}, Q^2)&\longrightarrow g(x, Q^2)\ \     \text{  with rate  } \ \   \frac{\alpha_s(Q^2)}{2\pi}P_{gq}(z),\nonumber
\\3)\ \   &q_i(\frac{x}{z}, Q^2)&\longrightarrow q_i(x, Q^2) \ \    \text{  with rate  } \ \   \frac{\alpha_s(Q^2)}{2\pi}P_{qq}(z),\nonumber
\\4)\ \   &g(\frac{x}{z}, Q^2)&\longrightarrow g(x, Q^2)  \ \    \text{  with rate  } \ \   \frac{\alpha_s(Q^2)}{2\pi}P_{gg}(z), \label{eq:processes}
\end{eqnarray}  
where $i'$ change from 1 to $n_F$, that $n_F$ is the number of  flavours of quark-antiquark pairs available to split the gluon into them and $i$ change from 1 to $2n_F$ (the number of quarks and antiquarks of all flavours). It should be noted that in the cases 1 and 2 (the birth-death processes), in addition to the longitudinal momentum variation of the parton, the type of parton can be also changed, while the cases 3 and 4 are the diffusion processes in the momentum space, i.e.,  only the longitudinal motion changes. It should be noted that according to the figure 1, the birth-death processes are accompanied by the emission of a quark, and the diffusion processes are accompanied by the emission of a gluon, which are not considered in the parton evolution chain (Markov chain).
\\Since the rate of the reaction diffusion processes of the partons inside the hadron, $\frac{\alpha_s(Q^2)}{2\pi}P_{aa'}(z)$,  is
 determined by using the phenomenological methods, the master equation for the  $Q^2$ evolution of  the $PDF$ is a phenomenological equation. As a result, the process of the parton $Q^2$ evolution in the leading order ($LO$) level of the $QCD$ is completely stochastic and depends only on the previous step. Therefore, it is the type of Markov processes in the statistical mechanics, and the master equation in the momentum space governs its $Q^2$ evolution. According to the $QCD$, the strong force between partons becomes very weak and asymptotically toward zero at short distances inside the hadron ($d<<10^{-15}m$). On the other hand, the high-energy experiments examine the structure of the hadron in a short time scales. Thus, the interactions between the partons can be ignored in comparison with their high energy interactions with the virtual photon \cite{20modarres}. In other words, with good approximation, the partons can be considered as the free particles inside the hadron i.e., the Feynman parton model \cite{20modarres}. Thus, the $Q^2$ evolution of partons can be studied in single-particle systems.
\\Based on mentioned points, in the study of the $PDF$, we can model the parton $Q^2$ evolution in the leading order level of the $QCD$ with the single-particle systems with the multi-species reaction diffusion processes, i.e., the birth-death processes (the cases 1 and 2 of equation (\ref{eq:processes})) and the diffusion processes (the cases 3 and 4 of equation (\ref{eq:processes})), on the finite one-dimensional continuous lattice in the momentum space (the longitudinal momentum axis). 
\\The master equation (\ref{eq:Master}) in the momentum space for the finite continuous stochastic variable of the fractional longitudinal momentum $x$ (corresponding to the site $x$ in the position space) at the hard scale $\mu$ (corresponding to the time scale $t$ in the position space) is rewritten as follows:
\begin{eqnarray}
\frac{\partial P_i(x, \mu)}{\partial \mu} =\sum_{j=1}^{p}\Bigg[\int_{0}^{1}dz \Bigg(P_{j}(\frac{x}{z}, \mu)\tau_{ij}(z, \mu)-
P_{i}(x, \mu)\tau_{ji}(z, \mu)\Bigg)\Bigg],
\label{eq:Master1}
\end{eqnarray}
where $P_i(x, \mu)$  is the probability of the finding the parton of type $i$ with the fraction $x$ of longitudinal momentum of the parent hadron inside the hadron by  probing photon with the energy scale $\mu$. Also, $\tau_{ij}(z, \mu)$ is the rate of transition $P_{j}(\frac{x}{z}, \mu)\rightarrow P_i(x, \mu)$, by the emission of a parton that is not considered in the parton evolution chain. It should be noted that this transition occurs by emitting parton $i$ (with the fraction of longitudinal momentum $x$) with  fraction $z$ of the longitudinal momentum of the parton $j$ (with the fraction of longitudinal momentum $\frac{x}{z}$).
Finally,  by replacing $xq_i(x, \mu)$ and $xg(x, \mu)$ instead of $P_i(x, \mu)$, $\frac{\alpha_s(\mu)}{2\pi}P_{ij}(z)$ instead of $\tau_{ij}(z, \mu)$ and  choosing the scale $\mu$ as $ Ln (\frac{Q^2}{{Q_0}^2})$  in the master equation (\ref{eq:Master1}), the equations for the $Q^2$ evolution of the $PDF$ by considering this fact that the strong force does not change the flavour of the quark (consequently, for example in the case of the quark distribution functions, $\sum_{j=1}^{p}$ in the master equation (\ref{eq:Master1}) is converted to $\sum_{j=1}^{2}$,  where $P_{1}(x, \mu)=xq_{i}(x, {Q}^2)$ and $P_{2}(x, \mu)=xg(x, {Q}^2)$), are obtained as follows:
\begin{eqnarray}
\frac{\partial q_i(x, Q^{2})}{\partial Ln Q^2} &=&\frac{\alpha_s({Q}^2)}{2\pi}\int_{0}^{1-\Delta}\frac{dz}{z}\Bigg[P_{qq}(z)q_i\left(\frac{x}{z}, {Q}^2 \right)+P_{qg}(z) g\left(\frac{x}{z}, {Q}^2 \right)\Bigg]\nonumber
 \\&-& \frac{\alpha_s({Q}^2)}{2\pi}\int_0^{1-\Delta}dz\Bigg[P_{qq}(z)q_i\left(x, {Q}^2 \right)+P_{gq}(z) q_i\left(x, {Q}^2 \right)\Bigg],\nonumber
 \\\frac{\partial g(x, Q^{2})}{\partial Ln Q^2} &=&\frac{\alpha_s({Q}^2)}{2\pi}\int_{0}^{1-\Delta}\frac{dz}{z}\Bigg[\sum_{i=1}^{2n_{F}}\Bigg( P_{gq}(z)q_i\left(\frac{x}{z}, {Q}^2 \right)\Bigg)+P_{gg}(z) g\left(\frac{x}{z}, {Q}^2 \right)\Bigg]\nonumber
 \\&-& \frac{\alpha_s({Q}^2)}{2\pi}\int_0^{1-\Delta}dz\Bigg[n_{F} P_{qg}(z)g\left(x, {Q}^2 \right)+P_{gg}(z) g\left(x, {Q}^2 \right)\Bigg].\label{eq:DGLAP1}
\end{eqnarray}
One then can recognize the positive terms as the sources of the $PDF$ and the negative terms as the sinks of it. Since $z = 1$ does not represent any reaction diffusion process, $\Delta$ is entered at the upper limit of the integrals.
The evolution equations obtained from stochastic modeling (the equation (\ref{eq:DGLAP1})) are the same as the $DGLAP$ evolution equations, but obtained much easier. 
 It should be noted that according to the evolution equation (\ref{eq:DGLAP1}), which is derived from (1) the non-equilibrium statistical mechanics, and (2) the results of reference \cite{Buccella} in relation to the boundary conditions governing this evolution equation, we hope that in the near future, by  solving methods of the master equation (for example by the coordinate Bethe ansatz method, see the reference \cite{ahmadi}, and the recursive method, see the reference \cite{olanj}), the new solutions for the $PDF$ are presented. 
\section{Results, discussions and conclusions}
It is shown that the process of the parton $Q^2$ evolution in the leading order level of the $QCD$ is completely stochastic and depends only on a previous step (Markov processes). The parton $Q^2$ evolution in the leading order level of the $QCD$ occurs by the emission of a parton, which changes the type and longitudinal momentum of the parton by the emission of a quark (the birth-death processes) or the longitudinal momentum of the parton by the emission of a gloun (the diffusion process). Therefore, the master equation in the momentum space governs its $Q^2$ evolution. It should be noted, based on the Feynman parton model \cite{20modarres}, the partons can be considered as the free particles inside the hadron. 
\\Therefore, the parton $Q^2$ evolution in the leading order level of the $QCD$ can be modeled by the single-particle systems with the multi-species reaction diffusion processes on the finite one-dimensional continuous lattice (the longitudinal momentum axis) in the momentum space. 
\\In conclusion, considering the outcome of the section $III$, we  extracted the $PDF$ evolution equations by using the stochastic modeling. These evolution equations are the same as the $DGLAP$ evolution equations, but they are obtained by much simpler mathematical procedure based on the non-equilibrium statistical mechanics and the theory of Markov processes. We hope in the near future, by solving methods of the master equation, the new solutions for the $PDF$ are presented. 

\begin{acknowledgements}
NO would like to acknowledge the University of Bu-Ali Sina for their
support. MM would also like to acknowledge the Research Council of
the University of Tehran for the grants provided for him.
\end{acknowledgements}

\newpage

\end{document}